\documentclass[fleqn,usenatbib]{mnras}

\usepackage{newtxtext,newtxmath}

\usepackage[T1]{fontenc}

\DeclareRobustCommand{\VAN}[3]{#2}
\let\VANthebibliography\thebibliography
\def\thebibliography{\DeclareRobustCommand{\VAN}[3]{##3}\VANthebibliography}

\usepackage{graphicx}	
\usepackage{amsmath}	

\title[Periodicity in repeating FRBs]{Are FRBs emitted from rotating magnetospheres? Searching for periodicity in polarized bursts.}

\author[Rajwade and Karastergiou]{
K M. Rajwade,$^{1}$\thanks{E-mail:kaustubh.rajwade@physics.ox.ac.uk}
A. Karastergiou,$^{1}$
\\
$^{1}$Astrophysics, University of Oxford, Denys Wilkinson Building, Keble road, Oxford OX1 3RH, UK\\
}

\date{Accepted XXX. Received YYY; in original form ZZZ}

\pubyear{\the\year{}}

\begin{document}
\label{firstpage}
\pagerange{\pageref{firstpage}--\pageref{lastpage}}
\maketitle

\begin{abstract}
One of the potential sources of repeating Fast Radio Bursts (FRBs) is a rotating magnetosphere of a compact object, as suggested by the similarities in the polarization properties of FRBs and radio pulsars.  Attempts to measure an underlying period in the times of arrival of repeating FRBs have nevertheless been unsuccessful. To explain this lack of observed periodicity, it is often suggested that the line of sight towards the source must be sampling active parts of the emitting magnetosphere throughout the rotation of the compact object, i.e. has a large duty cycle, as can be the case in a neutron star with near-aligned magnetic and rotation axes. This may lead to apparently aperiodic bursts, however the polarization angle of the bursts should be tied to the rotational phase from which they occur. This is true for radio pulsars. We therefore propose a new test to identify a possible stable rotation period under the assumptions above, based on a periodogram of the measured polarization angle timeseries for repeating FRBs. We show that this test is highly sensitive when the duty cycle is large, where standard time-of-arrival periodicity searches fail. Therefore, we can directly test the hypothesis of repeating FRBs of magnetospheric origin with a stable rotation period. Both positive and negative results of the test applied to FRB data will provide important information.

\end{abstract}




\section{Introduction}
Fast Radio Bursts (FRBs)~\citep{lorimer2007L} are characterized by their millisecond durations and their broadband, highly polarized emission. While a significant portion of detected FRBs appear as singular events, a subset exhibits repetition~\citep{spitler2016}. The distinction between repeating and non-repeating FRBs has profound implications for constraining their progenitor sources. These include, but are not limited to, highly magnetized neutron stars (magnetars), binary systems involving neutron stars or black holes, and even more exotic scenarios~\citep{lyutikov2020, zanazzi2020, beniamini2020}. Phenomenologically, FRBs share many observational traits with emission from radio pulsars.

To date, three repeating FRBs (hereafter Repeaters) have shown evidence of periodic activity cycles~\citep{Rajwade2020, r3periodic, pal2025}. This means that the repeaters emit bursts in a predictable time window. The activity cycles have been interpreted as a possible modulation of the observed emission from Repeaters due to an orbital period~\citep{wang2022, rajwade2023} or the precession of the magnetic axis of a compact object~\citep{zanazzi2020}. However, Repeaters show no periodicity in the times of arrival of the bursts themselves. Several studies have attempted to detect such periodicities, using time-domain analysis techniques~\citep{du2024, nimmo2023}, without success. It is worth noting that $\sim$220~ms quasi-periodic bursts were observed for the non-repeating FRB~20191221A, which could hint at the underlying spin of a putative compact object~\citep{chimefrb2022}.

~\cite{beniamini2024b} have recently provided an overview of observables from Repeaters originating from a rotating magnetosphere where the alignment between the rotation and magnetic axis, and the duty cycle are key concepts. The duty cycle is defined as the fraction of the rotational period in Repeaters for which the line of sight samples the active part of the magnetosphere, as in radio pulsars. Stochastic bursts from a near-aligned magnetosphere, where the duty cycle is large, will appear aperiodic in their times of arrival. By near-aligned, we mean specifically that the inclination angle is comparable to the opening angle of the radio emitting part of the magnetosphere. Here, we focus on the idea that in such aperiodic bursts, the angle of polarization is tied to the rotational phase of the star, as also suggested in \cite{lu2019}. This is evident in pulsars where the polarization position angle follows a particular trajectory with rotational phase~\citep[e.g.][]{rad1969}, and should be true for a coherent magnetospheric emission mechanism. It has also been used to obtain a correct timing solution for RRAT J1819-1458 \citep{J1819}\footnote{The third paragraph of Section 3.2 of \cite{J1819} is of particular relevance to this paper.}, where each narrow pulse carries the polarization signature of the rotational phase it is emitted from.
In this paper we propose a new method to find the rotational period of Repeaters by searching for periodicities in the polarization angle. We detail our assumptions and our methodology in Section 2. In Section 2.1 and 2.2, we test our hypothesis on simulated data and on data taken on single radio pulses from neutron stars. We discuss the implications of this method in Section 3 and summarize in Section 4.

\section{Methodology}

For this analysis, we make certain assumptions about the nature of Repeaters. The key assumption is that the origin of the bursts is magnetospheric~\citep{beniamini2024a, wadiasingh2020}. Most Repeaters are highly linearly polarized sources suggesting a connection between the location of the emitting region and the orientation of the magnetic field~\citep[e.g][]{beniamini2024b}. This means that the measured polarization position angle ($\psi$) must be a function of the rotational phase $\phi$ of the compact object such that,
\begin{equation}
\psi= \mathcal{G} \left(\phi \right),
\label{eq:psi}
\end{equation}
where, $\mathcal{G}(\phi)$ also depends on the geometry of the magnetic field on the surface of the compact object. For the case of dipolar fields as expected for neutron star, $\mathcal{G}(\phi)$ will depend on $\alpha$, the inclination angle of the magnetic axis from the rotation axis of the neutron star, and $\beta$, the impact parameter, defined as the smallest angle between the line of sight and the magnetic axis. For a rotating source, $\phi$ is a periodic function of time $t$,
\begin{equation}
    \phi(t)= \rm mod\left( 2\pi\nu t, 2\pi\right),
\label{eq:phi}
\end{equation}
where $\nu$ is the spin frequency. The combination of Eqs. \ref{eq:psi} and \ref{eq:phi} then means that $\mathcal{G}$ is a periodic function of $t$. Hence, every measured $\psi$ from a given Repeater should follow the same periodic relationship with time under the assumption that the period remains constant. Given a dataset of measurements of $\psi$ at different times $t$, we can establish this periodic relationship robustly using a periodogram~\citep[also see][]{lu2019}. 
Repeater bursts are irregular in time, hence we use a Lomb Scargle periodogram~\citep{Scargle1992} to identify periodicities in the timeseries of $\psi$. There are other methods like the Fast-Folding Algorithm (FFA)~\citep{staelin1969} that could also be used for this analysis. Any fast method that successfully retrieves periodicities from unevenly sampled data should do. We adhere to the Lomb-Scargle periodogram for which fast, readily available implementations exist.

For a given frequency $\nu$, the power in a Lomb-Scargle periodogram,
\begin{equation}
\begin{split}
P_{\rm LS}(\nu) = \frac{1}{2} \left(\sum_{n=1}^{N} \rm \psi(t_{n})~cos(2\pi\nu(t_{n} - \tau))\right)^{2}/\sum_{n=1}^{N} \rm cos^{2}(2\pi\nu(t_{n} - \tau)) \\
+ \frac{1}{2}\left( \sum_{n=1}^{N} \rm \psi(t_{n})~sin(2\pi\nu(t_{n} - \tau)) \right)^{2}/ \rm \sum_{n=1}^{N} sin^{2}(2\pi\nu(t_{n} - \tau)),
\end{split}
\end{equation}
where, 
\begin{equation}
\tau = \rm \frac{1}{4\pi\nu} tan^{-1}\left( \frac{\sum_{n=1}^{N} sin(4\pi\nu t_{n})}{\sum_{n=1}^{N} cos(4\pi\nu t_{n})}\right),
\end{equation}
and $N$ the number of measurements.
The value of $\tau$ is chosen such that it satisfies the time shift invariance~\citep[see][for more details]{vanderplas2018}. Here $\rm \psi(t_{n})$ is evaluated at the unevenly sampled times $\rm t_{n}$. 



\subsection{Simulations}

\begin{figure*}
	\includegraphics[width=\textwidth]{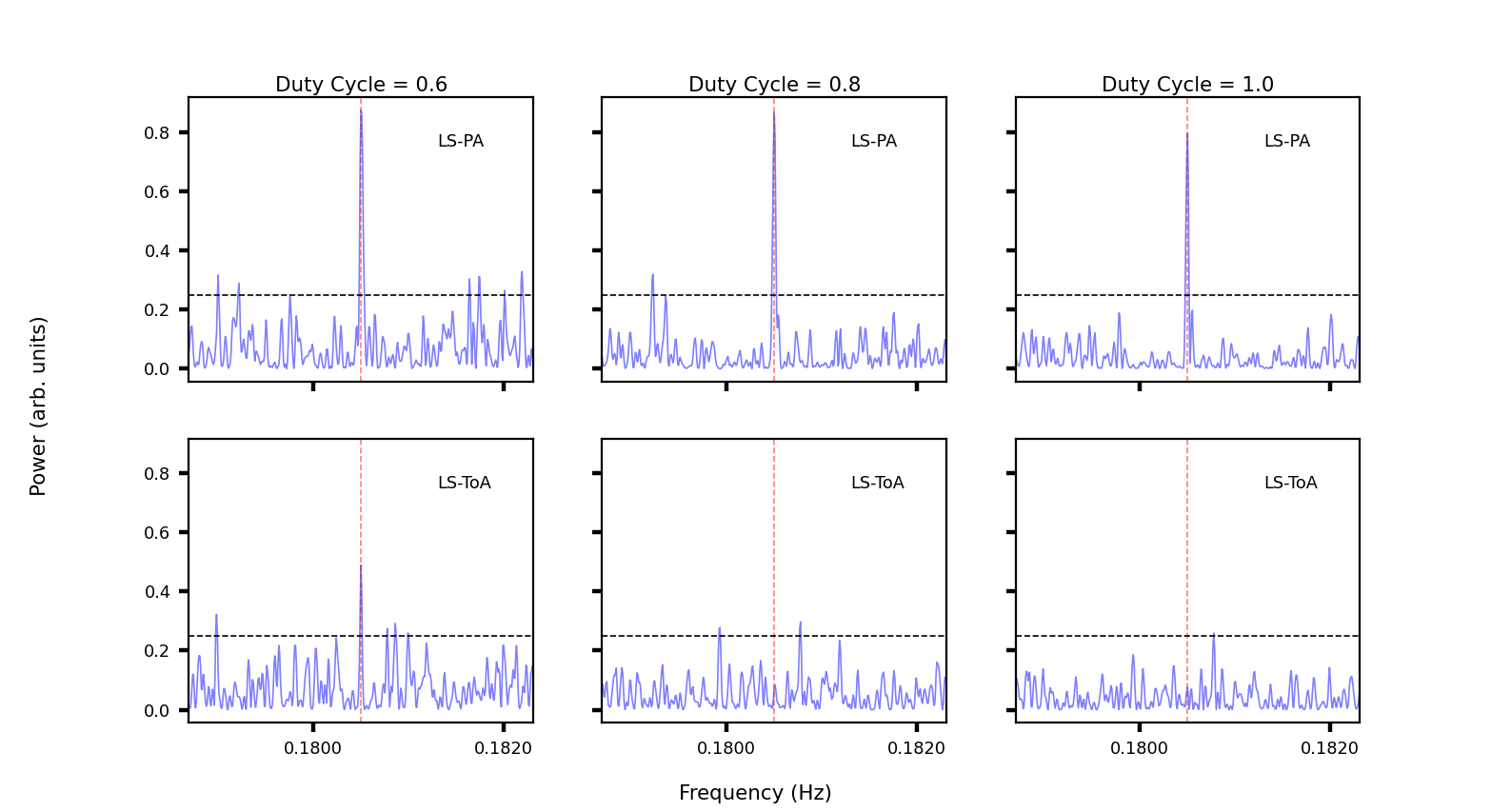}
    \caption{Lomb-Scargle periodogram of the polarization angle (top row) and ToAs (bottom row) from the simulations for a range duty cycles. The red dashed vertical line represents the rotation period of the star. The black dashed line corresponds to the false alarm probability of 0.00002$\%$.}
    \label{fig:sims}
\end{figure*}

To test our method, we set up a simulation whereby we sample $\psi$ for a near-aligned rotator. For our simulations, we assumed that $\mathcal{G}(\phi)$ is the rotating vector model, or RVM~\citep{rad1969}, where
\begin{equation}
\mathcal{G}(\phi) = \psi_{0}  + {\rm tan^{-1}}\left( \rm \frac{sin\alpha~sin(\phi - \phi_{0})}{cos\alpha~sin(\alpha + \beta) - sin\alpha~cos(\alpha + \beta)~cos(\phi - \phi_{0})}\right).
\label{eq:rvm}
\end{equation}
$\psi_{0}$ and $\phi_{0}$ are constants, and $\alpha=5^\circ$ and $\beta = 2^\circ$. The rotational period is set to $P=5.54$~s to mimic the period of a well-known radio-loud magnetar XTE~J1810-197~\citep{camilo2006}. The RVM would not be applicable for emission expected from multi-polar magnetic fields~\citep{yamasaki2022} but we consider RVM here as a test case, given most radio pulsars show RVM-like polarization profiles. It is essential to stress, however, that the results of this experiment do not require a particular dependence of $\mathcal{G}$ on $\phi$. We label a given rotation to be active for the rotations in which a burst occurs. The integer interval between consecutive active rotations is chosen from a random uniform distribution, from between 1 and $\texttt{maxinterval}=100$ rotations. For every active rotation, we generate the starting rotational phase $\phi_b$ of the burst from a uniform distribution between 0 and $2\pi$. We specify the duty cycle $\delta$ as the fraction of the period for which the line of sight samples the active region of the magnetosphere (bursts of much shorter duration can be emitted stochastically within the active region). If $\phi_b\leq 2\pi\delta $, the burst is detected, and we record the corresponding $\psi$ from Eq.~\ref{eq:rvm} and the timestamp for $\texttt{nsamp}=10$ consecutive samples of $\texttt{tsamp}=1$ ms each. Since typical bursts last for tens of milliseconds, one would expect to sample $\sim$10 points across a burst for 1~ms sampling interval which is normal for real-time surveys. We add an uncertainty to $\psi$ drawn from a normal distribution with mean zero and standard deviation of 0.1 radians. We run a Lomb-Scargle periodogram on the timeseries of $\psi$ and separately on the recorded timestamps of the detected bursts, only using the timestamp of the first of the 10 samples of each burst to mimic experimentally recorded times-of-arrival (TOAs) in FRB data.

Figure~\ref{fig:sims} shows the results of the above for 3 values of duty cycle, namely 0.6, 0.8, and 1. The periodogram applied to TOAs is unable to recover the rotational period for $\delta=0.8$ and $\delta=1$, but the periodogram applied to $\psi$ retains a high signal-to-noise peak at the expected spin frequency. This shows that while the periodicity search using burst TOAs does not result in any detection, using the values of $\psi$ will probe the underlying spin period if the bursts are emitted stochastically within the emission region of a nearly aligned rotator. The results of our simulations do not change if we chose the integer interval between consecutive active rotations from a Poisson distribution which is observed for wait-times of bursts from Repeaters~\citep{Cruces2021}. For the simulations, we also tried multiple different values for the period, and in each case, the method recovers the exact value.

\begin{figure*}
    \includegraphics[width=\textwidth]{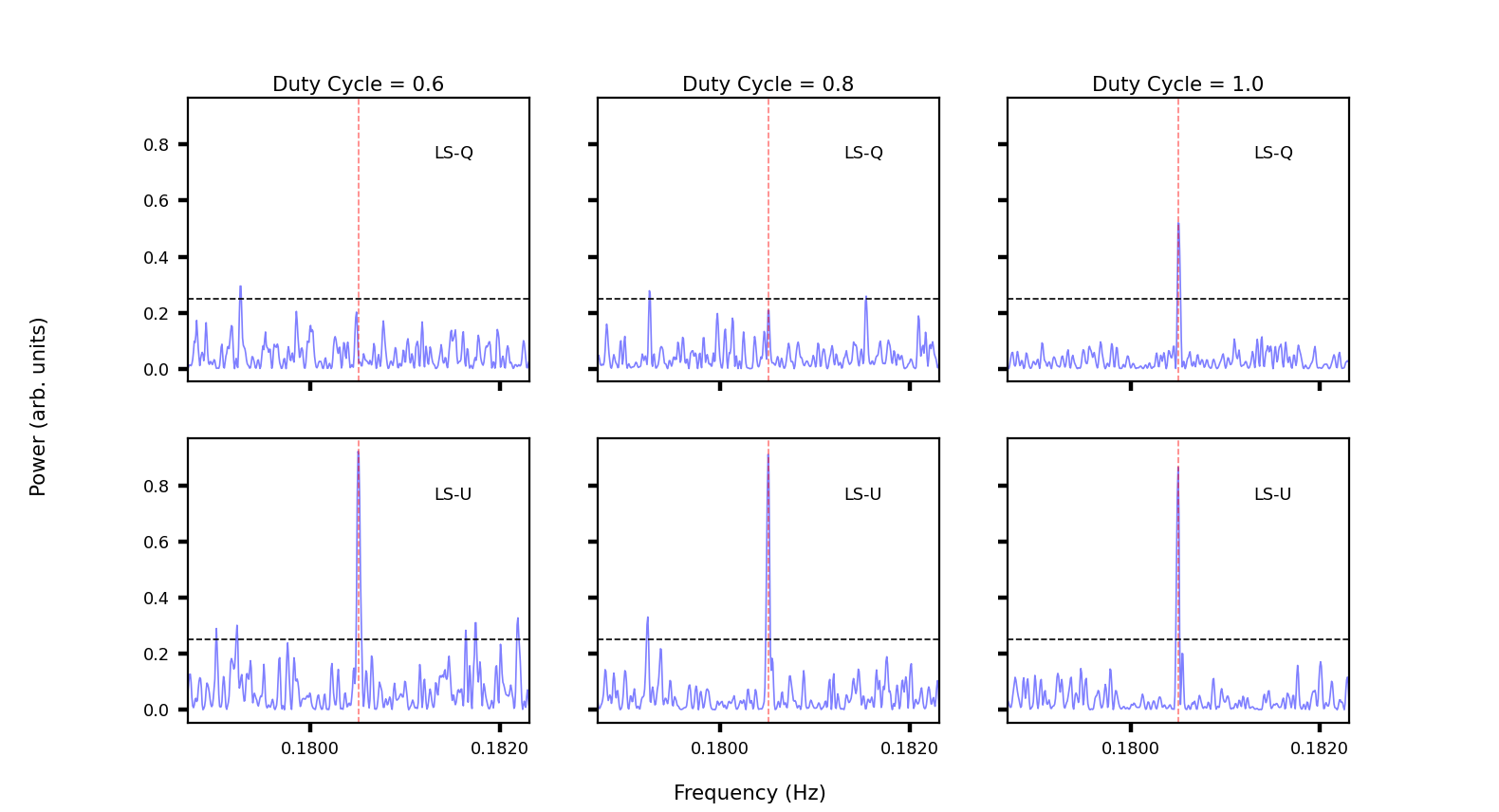}
    \caption{Lomb-Scargle periodogram of simulated normalized Stokes Q (top row) and Stokes U (bottom row) for different duty cycles. The red dashed vertical line shows the spin frequency and the dashed black line shows the threshold corresponding to a false alarm probability of 0.00002$\%$.}
    \label{fig:qu}
\end{figure*}

Next, we tested our methodology on radio pulses from well-known radio-emitting neutron stars. In our case, we chose the magnetar, XTE~J1810-197 which is a radio-loud neutron star with highly linearly polarized radio emission~\citep{kramer2007} and the $\psi$ values show a non-RVM like behaviour. We use single-pulse data taken by the MeerKAT radio telescope as part of the Thousand Pulsar Array programme~\citep{johnston2020}. We processed the full Stokes data with all the required polarization information and collected $\psi$ values where the linear polarization intensity greater is than 5$\sigma$ to ensure that we only sample signals from the pulsar. For our experiment, we randomly chose \texttt{n} pulses from our sample of \texttt{N} total pulses in the dataset. For each \texttt{n}, we noted each value of $\psi$ corresponding to the radio pulse and the associated time stamp to create a timeseries of $\psi$. Then we ran a LS periodogram on the resulting values. We detected a strong peak at the expected spin period of the pulsar using this method. However, we do note the low duty cycle of this source, which would therefore be easily detected in other TOA-based periodicity searches as well. Unlike the simulations above, this source does not strictly obey Eq. \ref{eq:rvm}. The method we are proposing here does not rely on a particular form of 
$\mathcal{G}(\phi)$.


\subsection{Discontinuities in the polarization position angle}
There is a key caveat to consider when using $\psi$ to search for periodicity, as it is defined inside an interval of $\pi$, as $\rm tan(2\psi) = Q/U$, where $\rm Q$ and $\rm U$ are the Stokes Q and Stokes U parameters. This interval is commonly defined as $[-\pi/2, \pi/2]$, and therefore the measured values of $\psi$ show discontinuities at the boundaries. Even for a given $\mathcal{G}(\phi)$, the uncertainty on $\psi$ means these discontinuities may occur at different values of $\phi$, which is not optimal for a periodicity search.

A solution to overcome this limitation is to directly compute a periodogram of the normalized Stokes Q and U values of the data, $\rm \frac{Q}{\sqrt{Q^2 + U^2}} = cos(2\psi)$ and $\frac{U}{\sqrt{Q^2 + U^2}}= sin(2\psi).$ Since the sine and cosine are always continuous over all angles, one can still recover the true period while avoiding the limitation due to the wraps. It is worth noting that the significance of peaks in the periodogram will depend on the actual values of $\psi$, and hence which of Stokes Q and U contains most of the linearly polarized power. Hence, we recommend that we add the periodogram from the normalized Stokes Q and U incoherently to maximize the chances of detection. Fig. \ref{fig:qu} shows an example of simulated data where the power is almost entirely in Stokes U.

\section{Discussion}

The sensitivity of this method depends on $\mathcal{G}$ not being a constant, which could arise in two scenarios. Firstly, if the duty cycle is small, the range of sampled phases $\phi$ is small, therefore the measured $\psi=\mathcal{G}(\phi)$ would tend to a constant value within uncertainties. It is worth noting that in this case, a periodicity should be detectable in the TOAs of the bursts. Secondly, for large duty cycles, the range of measured $\psi$ could be small depending on the emission geometry. Nevertheless, sampling $\mathcal{G}$ across a large range of $\phi$, maximizes the range of measured $\psi$ values. Therefore we expect this method to be most sensitive for large duty cycles, which is also the proposed explanation for aperiodic bursts from Repeaters, where TOA periodicity searches may fail~\citep{beniamini2024b}. 

Our method to find the intrinsic periodicity can be used for any set of bursts from Repeaters that show highly polarized emission where $\psi$ can be estimated accurately. This includes accounting for potentially variable Faraday rotation, and ensuring the polarization calibration provides position angles in a constant reference frame. The utility of the method is dictated by the correlation of the rotational phase and $\psi$, which then overcomes the limitation of the stochastic nature of the emission process within the active region of the magnetosphere. The sensitivity of the method is only dependent on the duty cycle. One needs to sample multiple rotations to detect the underlying period. If this condition is not satisfied within a single epoch of observation, it will not be picked up by this method. Hence, Repeaters with large periods (days) will be missed purely due to an observational bias. Some Repeaters have also exhibited $\psi$ values that vary with time i.e. do not show a consistently repeating pattern across the observations~\citep{niu2024}, making the search for a period with this method challenging. There are exceptional cases where the method may still fail to recover a period of rotation, even if one is fundamentally present. In the case of the rotating vector model, if the neutron star is a completely aligned rotator then regardless of the impact parameter, $\psi$ will be independent of the rotational phase, hence $\mathcal{G}={\rm const}$. Pulsar emission mechanisms in general require $\alpha$ to be non zero.

We note that this method can also be an effective tool to find periodicity for bursts that are seen across the entire rotational phase of neutron stars even when they are not necessarily aligned rotators in the sense of a small inclination angle $\alpha$, but where the radio beams are wide. An example is the sample of radio bursts seen from the magnetar SGR~J1935+2154 where the bursts have been seen across one full rotation of the star~\citep{kirsten2021a}. As long as the measured $\psi$ values are directly dependent on the orientation of the magnetic field of the region emitting the burst as, the methodology can be used to detect the intrinsic spin period. 

Ideally this method should be applied to datasets with a large number of bursts from a Repeater within a single observing epoch. A null result in such a search is also informative. It would suggest that the mechanism that is producing coherent emission in Repeaters is not related to the underlying physics that creates coherent radio emission in pulsars as a relationship between the rotational phase and polarization is a prediction from any such emission model. This would argue against a rotating magnetospheric origin for Repeaters. 

We caution using this method on datasets over several epochs spanning multiple days/weeks. If the period derivative of the neutron star producing the bursts is large, one might lose phase connection between subsequent epochs since $\psi$ is no longer a periodic function of $t$. Even a non-detection in such data yields some limits on the stability of the period. The TOAs of bursts will also be affected by timing noise, especially if the progenitors are magnetars~\citep{rajwade2022b}. The timing noise may dominate at those longer timescales resulting in a loss of phase connection between the $\psi$ values from different epochs.  

Another advantage of this method is that it can be used to detect the period of any radio-loud neutron star where a period has not be detected at radio wavelengths. Recent imaging surveys have led to the discovery of many highly polarized compact objects~\citep{callingham2023}. Prompt follow-up of such sources with time-domain radio observations has led to the discovery of pulsed radio emission \citep{sobey2022} confirming their neutron star nature but not all of the searches are successful. A reason for this could be that some of the sources have large duty cycles, as discussed here for Repeaters. Our method may identify the underlying spin-period of the potential pulsar if one is present.

\section{Summary}
In summary, we have presented a new methodology to identify periodicities in a set of polarized radio bursts emitted by Repeaters. The method assumes a relationship between the rotational phase and the polarization position angle of the radio emission~\citep{liu2025}, but not necessarily a rotating vector model. Testing this conjecture on simulated and real data shows that the method succeeds for large duty cycles, where TOA periodicity searches may fail. We strongly recommend that this methodology should be used on all existing datasets of Repeaters to identify the intrinsic spin period, especially multi-burst, single-epoch datasets. A null result from this experiment will have important implications on the interpretation of the origins of bursts from Repeaters. Given the information currently in the literature, we are optimistic that one of the existing high quality datasets from a large telescope will yield a positive result.

\section*{Acknowledgements}
The authors would like to thank the reviewer for their constructive comments on the manuscript. The authors would like to thank Harish Vedantham, Simon Johnston, and Alex Cooper for useful discussions that improved the clarity of the manuscript. The MeerKAT telescope is operated by the South African Radio
Astronomy Observatory, which is a facility of the National Research
Foundation, an agency of the Department of Science and Innovation. 
\section*{Data Availability}
The data and the code used in these analyses will be provided upon reasonable request to the authors.



\bibliographystyle{mnras}
\bibliography{example} 

\begin{thebibliography}{}
\makeatletter
\relax
\def\mn@urlcharsother{\let\do\@makeother \do\$\do\&\do\#\do\^\do\_\do\%\do\~}
\def\mn@doi{\begingroup\mn@urlcharsother \@ifnextchar [ {\mn@doi@} {\mn@doi@[]}}
\def\mn@doi@[#1]#2{\def\@tempa{#1}\ifx\@tempa\@empty \href {http://dx.doi.org/#2} {doi:#2}\else \href {http://dx.doi.org/#2} {#1}\fi \endgroup}
\def\mn@eprint#1#2{\mn@eprint@#1:#2::\@nil}
\def\mn@eprint@arXiv#1{\href {http://arxiv.org/abs/#1} {{\tt arXiv:#1}}}
\def\mn@eprint@dblp#1{\href {http://dblp.uni-trier.de/rec/bibtex/#1.xml} {dblp:#1}}
\def\mn@eprint@#1:#2:#3:#4\@nil{\def\@tempa {#1}\def\@tempb {#2}\def\@tempc {#3}\ifx \@tempc \@empty \let \@tempc \@tempb \let \@tempb \@tempa \fi \ifx \@tempb \@empty \def\@tempb {arXiv}\fi \@ifundefined {mn@eprint@\@tempb}{\@tempb:\@tempc}{\expandafter \expandafter \csname mn@eprint@\@tempb\endcsname \expandafter{\@tempc}}}

\bibitem[\protect\citeauthoryear{{Beniamini} \& {Kumar}}{{Beniamini} \& {Kumar}}{2024}]{beniamini2024b}
{Beniamini} P.,  {Kumar} P.,  2024, \mn@doi [arXiv e-prints] {10.48550/arXiv.2410.19043}, \href {https://ui.adsabs.harvard.edu/abs/2024arXiv241019043B} {p. arXiv:2410.19043}

\bibitem[\protect\citeauthoryear{{Beniamini}, {Wadiasingh}  \& {Metzger}}{{Beniamini} et~al.}{2020}]{beniamini2020}
{Beniamini} P.,  {Wadiasingh} Z.,   {Metzger} B.~D.,  2020, \mn@doi [\mnras] {10.1093/mnras/staa1783}, \href {https://ui.adsabs.harvard.edu/abs/2020MNRAS.496.3390B} {496, 3390}

\bibitem[\protect\citeauthoryear{{Beniamini}, {Wadiasingh}, {Trigg}, {Chirenti}, {Burns}, {Younes}, {Negro}  \& {Granot}}{{Beniamini} et~al.}{2024}]{beniamini2024a}
{Beniamini} P.,  {Wadiasingh} Z.,  {Trigg} A.,  {Chirenti} C.,  {Burns} E.,  {Younes} G.,  {Negro} M.,   {Granot} J.,  2024, \mn@doi [arXiv e-prints] {10.48550/arXiv.2411.16846}, \href {https://ui.adsabs.harvard.edu/abs/2024arXiv241116846B} {p. arXiv:2411.16846}

\bibitem[\protect\citeauthoryear{{Callingham} et~al.,}{{Callingham} et~al.}{2023}]{callingham2023}
{Callingham} J.~R.,  et~al., 2023, \mn@doi [\aap] {10.1051/0004-6361/202245567}, \href {https://ui.adsabs.harvard.edu/abs/2023A&A...670A.124C} {670, A124}

\bibitem[\protect\citeauthoryear{{Camilo}, {Ransom}, {Halpern}, {Reynolds}, {Helfand}, {Zimmerman}  \& {Sarkissian}}{{Camilo} et~al.}{2006}]{camilo2006}
{Camilo} F.,  {Ransom} S.~M.,  {Halpern} J.~P.,  {Reynolds} J.,  {Helfand} D.~J.,  {Zimmerman} N.,   {Sarkissian} J.,  2006, \mn@doi [\nat] {10.1038/nature04986}, \href {https://ui.adsabs.harvard.edu/abs/2006Natur.442..892C} {442, 892}

\bibitem[\protect\citeauthoryear{{Chime/Frb Collaboration} et~al.,}{{Chime/Frb Collaboration} et~al.}{2020}]{r3periodic}
{Chime/Frb Collaboration} et~al., 2020, \mn@doi [\nat] {10.1038/s41586-020-2398-2}, \href {https://ui.adsabs.harvard.edu/abs/2020Natur.582..351C} {582, 351}

\bibitem[\protect\citeauthoryear{{Chime/Frb Collaboration} Bridget~C. et~al.,}{{Chime/Frb Collaboration} et~al.}{2022}]{chimefrb2022}
{Chime/Frb Collaboration} Bridget~C. A.,  et~al., 2022, \mn@doi [\nat] {10.1038/s41586-022-04841-8}, \href {https://ui.adsabs.harvard.edu/abs/2022Natur.607..256C} {607, 256}

\bibitem[\protect\citeauthoryear{{Cruces} et~al.,}{{Cruces} et~al.}{2021}]{Cruces2021}
{Cruces} M.,  et~al., 2021, \mn@doi [\mnras] {10.1093/mnras/staa3223}, \href {https://ui.adsabs.harvard.edu/abs/2021MNRAS.500..448C} {500, 448}

\bibitem[\protect\citeauthoryear{{Du}, {Huang}, {Zhang}, {Rodin}, {Fedorova}, {Kurban}  \& {Li}}{{Du} et~al.}{2024}]{du2024}
{Du} C.,  {Huang} Y.-F.,  {Zhang} Z.-B.,  {Rodin} A.,  {Fedorova} V.,  {Kurban} A.,   {Li} D.,  2024, \mn@doi [\apj] {10.3847/1538-4357/ad8cd5}, \href {https://ui.adsabs.harvard.edu/abs/2024ApJ...977..129D} {977, 129}

\bibitem[\protect\citeauthoryear{{Johnston} et~al.,}{{Johnston} et~al.}{2020}]{johnston2020}
{Johnston} S.,  et~al., 2020, \mn@doi [\mnras] {10.1093/mnras/staa516}, \href {https://ui.adsabs.harvard.edu/abs/2020MNRAS.493.3608J} {493, 3608}

\bibitem[\protect\citeauthoryear{{Karastergiou}, {Hotan}, {van Straten}, {McLaughlin}  \& {Ord}}{{Karastergiou} et~al.}{2009}]{J1819}
{Karastergiou} A.,  {Hotan} A.~W.,  {van Straten} W.,  {McLaughlin} M.~A.,   {Ord} S.~M.,  2009, \mn@doi [\mnras] {10.1111/j.1745-3933.2009.00671.x}, \href {https://ui.adsabs.harvard.edu/abs/2009MNRAS.396L..95K} {396, L95}

\bibitem[\protect\citeauthoryear{{Kirsten}, {Snelders}, {Jenkins}, {Nimmo}, {van den Eijnden}, {Hessels}, {Gawro{\'n}ski}  \& {Yang}}{{Kirsten} et~al.}{2021}]{kirsten2021a}
{Kirsten} F.,  {Snelders} M.~P.,  {Jenkins} M.,  {Nimmo} K.,  {van den Eijnden} J.,  {Hessels} J.~W.~T.,  {Gawro{\'n}ski} M.~P.,   {Yang} J.,  2021, \mn@doi [Nature Astronomy] {10.1038/s41550-020-01246-3}, \href {https://ui.adsabs.harvard.edu/abs/2021NatAs...5..414K} {5, 414}

\bibitem[\protect\citeauthoryear{{Kramer}, {Stappers}, {Jessner}, {Lyne}  \& {Jordan}}{{Kramer} et~al.}{2007}]{kramer2007}
{Kramer} M.,  {Stappers} B.~W.,  {Jessner} A.,  {Lyne} A.~G.,   {Jordan} C.~A.,  2007, \mn@doi [\mnras] {10.1111/j.1365-2966.2007.11622.x}, \href {https://ui.adsabs.harvard.edu/abs/2007MNRAS.377..107K} {377, 107}

\bibitem[\protect\citeauthoryear{{Liu} et~al.,}{{Liu} et~al.}{2025}]{liu2025}
{Liu} X.,  et~al., 2025, \mn@doi [arXiv e-prints] {10.48550/arXiv.2504.00391}, \href {https://ui.adsabs.harvard.edu/abs/2025arXiv250400391L} {p. arXiv:2504.00391}

\bibitem[\protect\citeauthoryear{{Lorimer}, {Bailes}, {McLaughlin}, {Narkevic}  \& {Crawford}}{{Lorimer} et~al.}{2007}]{lorimer2007L}
{Lorimer} D.~R.,  {Bailes} M.,  {McLaughlin} M.~A.,  {Narkevic} D.~J.,   {Crawford} F.,  2007, \mn@doi [Science] {10.1126/science.1147532}, \href {https://ui.adsabs.harvard.edu/abs/2007Sci...318..777L} {318, 777}

\bibitem[\protect\citeauthoryear{{Lu}, {Kumar}  \& {Narayan}}{{Lu} et~al.}{2019}]{lu2019}
{Lu} W.,  {Kumar} P.,   {Narayan} R.,  2019, \mn@doi [\mnras] {10.1093/mnras/sty2829}, \href {https://ui.adsabs.harvard.edu/abs/2019MNRAS.483..359L} {483, 359}

\bibitem[\protect\citeauthoryear{{Lyutikov}, {Barkov}  \& {Giannios}}{{Lyutikov} et~al.}{2020}]{lyutikov2020}
{Lyutikov} M.,  {Barkov} M.~V.,   {Giannios} D.,  2020, \mn@doi [\apjl] {10.3847/2041-8213/ab87a4}, \href {https://ui.adsabs.harvard.edu/abs/2020ApJ...893L..39L} {893, L39}

\bibitem[\protect\citeauthoryear{{Nimmo} et~al.,}{{Nimmo} et~al.}{2023}]{nimmo2023}
{Nimmo} K.,  et~al., 2023, \mn@doi [\mnras] {10.1093/mnras/stad269}, \href {https://ui.adsabs.harvard.edu/abs/2023MNRAS.520.2281N} {520, 2281}

\bibitem[\protect\citeauthoryear{{Niu} et~al.,}{{Niu} et~al.}{2024}]{niu2024}
{Niu} J.~R.,  et~al., 2024, \mn@doi [\apjl] {10.3847/2041-8213/ad7023}, \href {https://ui.adsabs.harvard.edu/abs/2024ApJ...972L..20N} {972, L20}

\bibitem[\protect\citeauthoryear{{Pal}}{{Pal}}{2025}]{pal2025}
{Pal} A.,  2025, \mn@doi [arXiv e-prints] {10.48550/arXiv.2502.11215}, \href {https://ui.adsabs.harvard.edu/abs/2025arXiv250211215P} {p. arXiv:2502.11215}

\bibitem[\protect\citeauthoryear{{Radhakrishnan} \& {Cooke}}{{Radhakrishnan} \& {Cooke}}{1969}]{rad1969}
{Radhakrishnan} V.,  {Cooke} D.~J.,  1969, \aplett, \href {https://ui.adsabs.harvard.edu/abs/1969ApL.....3..225R} {3, 225}

\bibitem[\protect\citeauthoryear{{Rajwade} \& {van den Eijnden}}{{Rajwade} \& {van den Eijnden}}{2023}]{rajwade2023}
{Rajwade} K.~M.,  {van den Eijnden} J.,  2023, \mn@doi [\aap] {10.1051/0004-6361/202245468}, \href {https://ui.adsabs.harvard.edu/abs/2023A&A...673A.136R} {673, A136}

\bibitem[\protect\citeauthoryear{{Rajwade} et~al.,}{{Rajwade} et~al.}{2020}]{Rajwade2020}
{Rajwade} K.~M.,  et~al., 2020, \mn@doi [\mnras] {10.1093/mnras/staa1237}, \href {https://ui.adsabs.harvard.edu/abs/2020MNRAS.495.3551R} {495, 3551}

\bibitem[\protect\citeauthoryear{{Rajwade} et~al.,}{{Rajwade} et~al.}{2022}]{rajwade2022b}
{Rajwade} K.~M.,  et~al., 2022, \mn@doi [\mnras] {10.1093/mnras/stac446}, \href {https://ui.adsabs.harvard.edu/abs/2022MNRAS.512.1687R} {512, 1687}

\bibitem[\protect\citeauthoryear{{Scargle}}{{Scargle}}{1992}]{Scargle1992}
{Scargle} J.~D.,  1992, in {Feigelson} E.~D.,  {Babu} G.~J.,  eds, Statistical Challenges in Modern Astronomy. pp 411--436, \mn@doi{10.1007/978-1-4613-9290-3_47}

\bibitem[\protect\citeauthoryear{{Sobey} et~al.,}{{Sobey} et~al.}{2022}]{sobey2022}
{Sobey} C.,  et~al., 2022, \mn@doi [\aap] {10.1051/0004-6361/202142636}, \href {https://ui.adsabs.harvard.edu/abs/2022A&A...661A..87S} {661, A87}

\bibitem[\protect\citeauthoryear{{Spitler} et~al.,}{{Spitler} et~al.}{2016}]{spitler2016}
{Spitler} L.~G.,  et~al., 2016, \mn@doi [\nat] {10.1038/nature17168}, \href {https://ui.adsabs.harvard.edu/abs/2016Natur.531..202S} {531, 202}

\bibitem[\protect\citeauthoryear{{Staelin}}{{Staelin}}{1969}]{staelin1969}
{Staelin} D.~H.,  1969, \mn@doi [IEEE Proceedings] {10.1109/PROC.1969.7051}, \href {https://ui.adsabs.harvard.edu/abs/1969IEEEP..57..724S} {57, 724}

\bibitem[\protect\citeauthoryear{{VanderPlas}}{{VanderPlas}}{2018}]{vanderplas2018}
{VanderPlas} J.~T.,  2018, \mn@doi [\apjs] {10.3847/1538-4365/aab766}, \href {https://ui.adsabs.harvard.edu/abs/2018ApJS..236...16V} {236, 16}

\bibitem[\protect\citeauthoryear{{Wadiasingh}, {Beniamini}, {Timokhin}, {Baring}, {van der Horst}, {Harding}  \& {Kazanas}}{{Wadiasingh} et~al.}{2020}]{wadiasingh2020}
{Wadiasingh} Z.,  {Beniamini} P.,  {Timokhin} A.,  {Baring} M.~G.,  {van der Horst} A.~J.,  {Harding} A.~K.,   {Kazanas} D.,  2020, \mn@doi [\apj] {10.3847/1538-4357/ab6d69}, \href {https://ui.adsabs.harvard.edu/abs/2020ApJ...891...82W} {891, 82}

\bibitem[\protect\citeauthoryear{{Wang}, {Zhang}, {Dai}  \& {Cheng}}{{Wang} et~al.}{2022}]{wang2022}
{Wang} F.~Y.,  {Zhang} G.~Q.,  {Dai} Z.~G.,   {Cheng} K.~S.,  2022, \mn@doi [Nature Communications] {10.1038/s41467-022-31923-y}, \href {https://ui.adsabs.harvard.edu/abs/2022NatCo..13.4382W} {13, 4382}

\bibitem[\protect\citeauthoryear{{Yamasaki}, {Ek{\c{s}}i}  \& {G{\"o}{\u{g}}{\"u}{\c{s}}}}{{Yamasaki} et~al.}{2022}]{yamasaki2022}
{Yamasaki} S.,  {Ek{\c{s}}i} K.~Y.,   {G{\"o}{\u{g}}{\"u}{\c{s}}} E.,  2022, \mn@doi [\mnras] {10.1093/mnras/stac699}, \href {https://ui.adsabs.harvard.edu/abs/2022MNRAS.512.3189Y} {512, 3189}

\bibitem[\protect\citeauthoryear{{Zanazzi} \& {Lai}}{{Zanazzi} \& {Lai}}{2020}]{zanazzi2020}
{Zanazzi} J.~J.,  {Lai} D.,  2020, \mn@doi [\apjl] {10.3847/2041-8213/ab7cdd}, \href {https://ui.adsabs.harvard.edu/abs/2020ApJ...892L..15Z} {892, L15}

\makeatother
\end{thebibliography}





\bsp	
\label{lastpage}
\end{document}